\documentclass[twocolumn,prl,aps,nobalancelastpage,amsfonts,floatfix]{revtex4}

\usepackage{graphicx}
\usepackage{bm}
\begin{document}

\title{Evidence for nodal superconductivity in LaFePO}

\author{J.D. Fletcher,$^1$ A. Serafin,$^1$  L. Malone,$^1$ J. Analytis,$^2$ J-H Chu,$^2$ A.S. Erickson,$^2$ I.R. Fisher,$^2$ and A. Carrington,$^1$}
\affiliation{$^1$ H.H. Wills Physics Laboratory, University of Bristol, Tyndall Avenue, BS8 1TL, United Kingdom.}
\affiliation{$^2$Geballe Laboratory for Advanced Materials and Department of Applied Physics, Stanford University,
Stanford, California 94305-4045}

\date{\today}

\begin{abstract}
In several iron-arsenide superconductors there is strong evidence for a fully gapped superconducting state consistent
with either a conventional $s$-wave symmetry or an unusual $s_\pm$ state where there the gap changes sign between the
electron and hole Fermi surface sheets. Here we report measurements of the penetration depth $\lambda(T)$ in very clean
samples of the related iron-phosphide superconductor, LaFePO,  at temperatures down to $\sim$ 100 mK. We find that
$\lambda(T)$ varies almost perfectly linearly with $T$ strongly suggesting the presence of gap nodes in this compound.
Taken together with other data, this suggests the gap function may not be generic to all pnictide superconductors.
\end{abstract}
\maketitle

A key question in understanding the physics of the recently discovered iron pnictide superconductors is the origin of
the pairing interaction. Measurements of the symmetry and anisotropy of the superconducting energy gap are very useful
for helping to decide between competing theories. Although several experiment designed to deduce the gap anisotropy
have been performed a consensus has not yet been reached.

Several experiments indicate that the gap is finite at all points on the Fermi surface, however, others have suggested
the presence of pronounced gap anisotropy or nodes. The presence of a fully gapped state is supported by point-contact
tunneling spectroscopy measurements of SmFeAsO$_x$F$_y$ (Sm-1111) \cite{TYChenNature08,DagheroCMDec08,GonnelliCM08},
penetration depth measurements of Pr-1111 \cite{Hashimoto08a}, Sm-1111 \cite{Malone08}, Nd-1111 \cite{MartinCM08}, and
Ba$_x$K$_y$Fe$_2$As$_2$ (Ba-122) \cite{Hashimoto08b} and angle resolved photoemission measurements of Ba-122
\cite{DingH08,Zhao_Lin08,Nakayama08,Evtushinksy08}. Several of these measurements show evidence of two distinct gaps.
On the other hand, some experiments suggest the presence of low energy excitations which could be indicative of nodes.
These include nuclear magnetic resonance measurements of La-1111, \cite{Nakai08,Grafe08} and penetration depth
measurements of Ba-122 \cite{Gordon08}.

Theoretically, there is considerable debate about the nature of the gap anisotropy and symmetry. Band-structure
calculations have indicated that both the 1111 and 122 materials have multiple quasi-two-dimensional sheets of
Fermi-surface, hole-like close to the zone center and electron-like at the zone corner.  At least for the undoped
parent compounds these sheets are close to a nesting instability which can drive the system towards antiferromagnetism
\cite{MazinspmPRL}. Several papers have argued that spin-fluctuations mediate the electron pairing and Mazin \textit{et
al.} \cite{MazinspmPRL} argued this favors an $s_\pm$ paring state where both the electron and hole Fermi surfaces are
fully gapped but with gap-functions which are $\pi$ out of phase.  Others however, have come to different conclusions.
Most recently, Graser \textit{et al.} \cite{Graser08120343} have calculated that the two lowest energy gap functions,
one with $s$ symmetry and one with $d$ symmetry both have nodes on one or more of the Fermi surface sheets and are
nearly degenerate.

The iron phosphide superconductor LaFePO, is isostructural with LaFeAsO (La-1111). Its relatively low $T_c \sim$ 6~K
\cite{Kamihara2006} has been linked to the fact that the Fe-P bond angles depart substantially from those of a regular
tetrahedron \cite{CHLee08}. Unlike the corresponding As-based compound, nominally undoped LaFePO is non-magnetic and
superconducting \cite{Analytis2008}.  The electronic structure of LaFePO has recently been explored in some detail by
de Haas-van Alphen (dHvA) measurements \cite{Coldea2008}, which confirm the band-structure predictions of almost nested
electron and hole pockets,  a feature expected to be common to all Fe-pnictides not subject to antiferromagnetic Fermi
surface reconstruction. The detailed knowledge of the electronic structure and availability of superconducting samples
with very low levels of disorder makes LaFePO an ideal material in which to study order parameter symmetry.  In this
letter, we report measurements of the London penetration depth of LaFePO down to very low temperature $T<0.02~T_c$. Our
data show strong evidence for the presence of line-nodes in this compound.

\begin{figure}
\center
\includegraphics*[width=7cm]{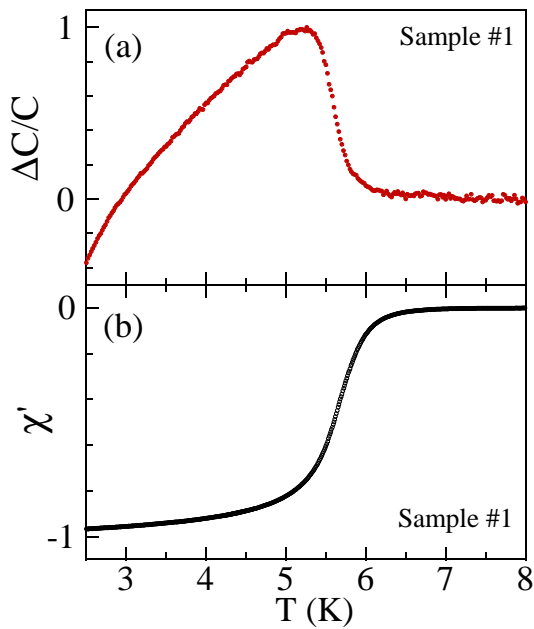}
\caption{(a) Heat capacity anomaly in a LaFePO single crystal. $\Delta C$ is the difference between the heat capacity
data in zero field and for $B=0.47~T$ (where $T_c<$2~K). (b) RF susceptibility of the same sample close to $T_c$.}
\label{tcfig}
\end{figure}

Our samples were grown via a flux method which produces plate-like single crystals with typical dimensions 0.15\,mm in
the basal plane and $<$0.03\,mm in the interlayer direction \cite{Analytis2008}. The samples measured in this study are
from the same batch as those used for dHvA measurements where the mean free path was estimated to be $\gtrsim$
1000\,\AA~for the electron sheets and $\sim$ 500\,\AA~for the hole sheets \cite{Coldea2008}.  Heat capacity, measured
using a modulated temperature technique, on the same single crystals used for the penetration depth measurements
confirmed the bulk nature of the superconductivity. The data shown in Fig.\ \ref{tcfig} shows a sharp jump in $C$ at
$T_c$ with the midpoint at $\simeq $5.6\,K with width $\sim$0.25\,K.

Measurements of the temperature dependence of the London penetration depth were performed with a high resolution
susceptometer based on a self-resonant tunnel diode circuit which was mounted in a dilution refrigerator. The circuit
operates at $\simeq$ 14~MHz with an extremely small probe field ($H_{ac}<10$mOe) so that the sample is always in the
Meissner state \cite{CarringtonGKG99}. Changes in the resonant frequency are directly proportional to changes in the
magnetic penetration depth as the temperature of the sample is varied.  The calibration factor is determined from the
geometry of the sample, and the total perturbation to the resonant frequency due to the sample, found by withdrawing
the sample from the coil at low temperature \cite{ProzorovGCA00}.  The sample is mounted on a sapphire rod, the other
end of which is glued to a copper block on which the RuO$_2$ thermometer is mounted.   The sample and rod are placed
inside a solenoid which forms part of the resonant tank circuit.  The position of the sample within the solenoid could
be varied in-situ allowing us to vary the RF field seen by the sample and hence check for RF heating which in-principle
could cause the thermometer and sample to be out of equilibrium. The sapphire sample holder is of very high purity and
has a very small paramagnetic background signal which was difficult to determine precisely and has not been subtracted.
It corresponds to at most 20\,\AA~ at the lowest temperature. Susceptibility measurements in this apparatus are shown
in Fig.\ \ref{tcfig} in the region near $T_c$. The mid point of the transition occurs at 5.6/,K consistent with the
heat capacity measurement on the same sample. Other samples measured had $T_c$ values in the range 5.4-5.9\,K.

\begin{figure}
\center
\includegraphics*[width=7.5cm]{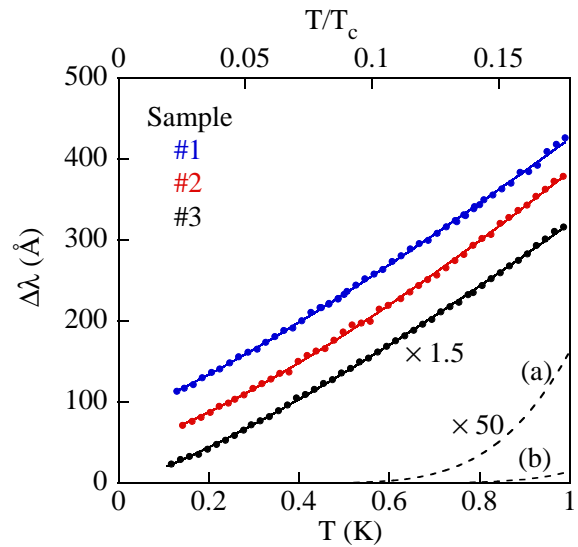}
\caption{Temperature dependence of the in-plane penetration depth, $\Delta\lambda_a$ in three single crystals of
LaFePO. Data for sample \#3 has been multiplied by 1.5. The curves are offset for clarity. Solid lines are power-law
fits giving an exponent of $1.2\pm 0.1$. Curve (a) is an extrapolation of the two gap behavior found in Sm-1111
\cite{Malone08}. Curve (b) is the temperature dependence expected from weak coupling BCS behavior (i.e. $\Delta =
1.76~T_c$). We have used $\lambda(0) \sim 2500$\AA~ in both cases and multiplied both curves by 50 to make them visible
on this scale.}\label{rawlamfig}
\end{figure}

Fig.\ \ref{rawlamfig} shows the temperature dependence of the in-plane penetration depth, $\Delta\lambda_a(T)$ in three
samples, all measured with the ac field perpendicular to the conducting planes so only in-plane currents are induced.
The temperature dependence is very similar in all three samples and is approximately linear. A variable power law fit,
i.e. $\Delta\lambda(T) \propto T^n$ for $T<1K$ gives $n=1.2\pm0.1$. The close to linear temperature dependence of
$\lambda$ is strongly indicative of nodes in the order parameter.

The temperature dependence of $\lambda$ arises from the depletion of the screening superfluid by thermal excitations.
The superconducting energy gap, $\Delta_k$, dictates the spectrum of these excitations; if $\Delta_k$ is finite for all
momentum wavevectors, an exponential temperature dependence of $\lambda(T)$ is observed. By contrast, in a
superconductor with an unconventional pairing symmetry the order parameter passes through zero for some wavevectors,
leading to the presence of low energy excitations down to low temperatures and a power-law temperature dependence of
the penetration depth. In a clean superconductor with line nodes on a quasi 2D fermi surface the angular averaged
density of states has a linear energy dependence, leading to a linear behavior of $\lambda(T)$ at low temperature.

The linear temperature dependence in LaFePO is in marked contrast with that observed in conventional superconductors,
or those displaying anisotropic $s$-wave superconductivity, such as MgB$_2$ or NbSe$_2$ \cite{ManzanoCHLYT02,
fletcher07}, or the As-based 1111 compounds described in the introduction. In an $s$-wave multiband superconductor the
low temperature behavior is dominated by the smallest gap on the Fermi surface. We show in Fig.\ \ref{rawlamfig} the
expected temperature dependence from gaps of 1.76~$T_c$ (the BCS weak coupling value) and 1.0~$T_c$ (similar to that
found in Sm-1111 \cite{Malone08}). These show a clear exponential saturation which is not visible in our data. We find
that if there is a small residual gap it at least 30 times smaller than the gap maximum. The presence of nodes is
therefore much more likely.

\begin{figure}
\center
\includegraphics*[width=7.5cm]{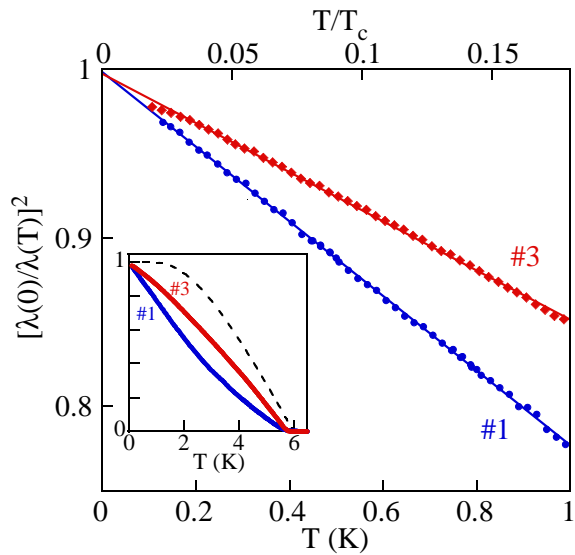}
\caption{Low temperature normalized superfluid density for samples \#1 and \#3 using $\lambda(0)=2500$\AA. Solid lines
are linear fits to the data. Inset shows the superfluid density over the full temperature range. The dashed line shows
the behavior found for Sm-1111 \cite{Malone08}.}\label{figrhos}
\end{figure}

The absolute values of $\Delta\lambda(T)$ are somewhat variable between samples.  As shown in Fig.\ \ref{rawlamfig},
$d\lambda/dT  \simeq 350$\,\AA/Hz in samples \#1 and \#2 and smaller value of 220\,\AA/Hz in sample \#3. This
discrepancy in slope is likely to be due to uncertainties in the calibration factors relating the measured frequency
shifts to $\Delta\lambda(T)$ arising from surface roughness.  LaFePO has a mica-like morphology and samples \#1 and \#2
had visibly more flaky edges than sample\#3, which had mirror like surface on all faces (a similar effect was observed
in NbSe$_2$ \cite{fletcher07}).  Alternatively, it could result from intrinsic differences in the superfluid density as
a function of doping i.e., slight changes in the oxygen stoichiometry (the samples had slightly different $T_c$).

In the usual weak-coupling theory for superconductors with line nodes, the superfluid density $\rho_s =
[\lambda(0)/\lambda(T)]^2$ varies linearly over a much wider range of $T$ than $\lambda(T)$ (corrections to the former
are $\mathcal{O}T^3$ whereas to the later they are of $\mathcal{O}T^2$). Muon spin relaxation measurements of
polycrystalline samples \cite{Uemura08} report a depolarisation rate of 1.2$\mu$m$^{-1}$, which corresponds to a
penetration depth of $\lambda(0) \simeq 2400$\,\AA~(we note that these authors indicate some uncertainty in this result
due to a comparable anomalous contribution to the depolarisation rate in those samples). Measurements in the related
materials La-1111 \cite{Khasanov08051923} and Sm-1111 \cite{HLuetkens08043115} span the range 2000-3500~\AA. Choosing a
representative value of $\lambda(0)$=2500\,\AA~  we calculate the superfluid density as shown in Fig.\ \ref{figrhos}.
The figure shows that despite the uncertainties in the ratio $\Delta\lambda(T)/\lambda(0)$ the superfluid density
varies almost perfectly linearly with $T$ from below $\sim$ 2\,K to around 100 mK.  This key result is robust to the
assumed value of $\lambda(0)$, so there is strong evidence for nodal excitations in the data regardless of this choice.
In a nodal superconductor the low temperature behavior is dominated by the excitations near the gap nodes. For a
superconductor with line-nodes $\lambda(T)$ is linear, with a slope determined by the rate at which the gap grows away
from the nodes. In the simple $d$-wave case with $\Delta_k = \Delta_0 \cos(2\theta)$, $\Delta\lambda = \ln 2 \lambda(0)
k_B T/\Delta_0$. For the weak coupling value $\Delta_0 \simeq 2.14~k_BT_c$ this predicts a value of around 100-200
\AA/K (for $\lambda(0)$ in the range quoted above), somewhat smaller than the slope observed experimentally in sample
\#1 and \#2, but close to the value seen in sample \#3. As this calculation relies directly on $\lambda(0)$, better
measurements of this parameter are required to make any precise comparisons to this or other candidate gap functions.

\begin{figure}
\center
\includegraphics*[width=7.5cm]{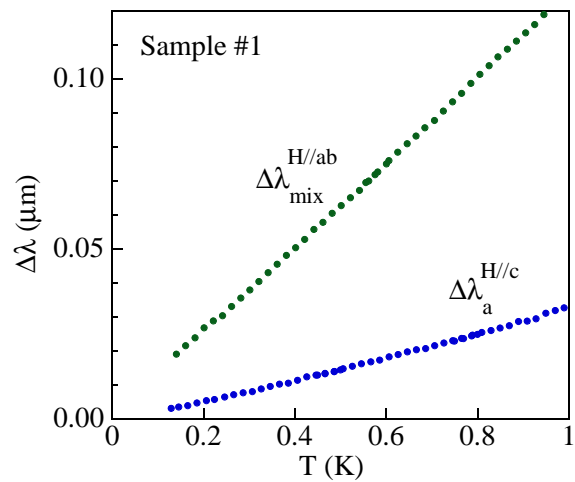}
\caption{Temperature dependence of the penetration depth, $\Delta\lambda_{\rm mix}$, measured with $H\parallel ab$.
Data for the same crystal with $H\parallel c$ are also shown for comparison.}\label{figlamc}
\end{figure}

Differences in the temperature dependence of $\lambda_c$ and $\lambda_a$ can sometimes reveal information about gap
symmetry. For example, in multi-band systems the $c$-axis response can also be different due to the presence of sheets
with differing anisotropy and different gaps \cite{Fletcher2005,fletcher07}. For sample \#1 measurements were also
performed with the field parallel to the conducting planes. This induces a combination of in-plane and inter-plane
currents and thus probes a mixture of $\Delta\lambda_a$ and $\Delta\lambda_c$. Sample \#1 has a ratio of thickness to
width of about $t/w \simeq 7$ which mixes in a substantial component of $\lambda_c$. We denote this mixture
$\Delta\lambda_{\rm mix}$. This data is shown in Fig.\ \ref{figlamc} along with data from H$\|c$ for comparison. The
changes in $\lambda_{\rm mix}$ are larger, with $d\lambda_{\rm mix}/dT = 1300$ \AA/Hz. Although the present crystals
are not of sufficiently regular geometry to isolate the changes in $\lambda_c$ precisely, the data are consistent with
$\lambda_c(T)$ also having a linear temperature dependence but with a much larger slope due to substantial electronic
anisotropy. Measurements of the upper critical field at low temperature indicate an anisotropy
$H_{c2}^{\|a}/H_{c2}^{\|c} \simeq 11$ consistent with the quasi-2D character of the Fermi surface \cite{Coldea2008}.
When line nodes are present on a simply warped quasi-2D Fermi surface, as found in LaFePO, $\lambda_c(T)/\lambda_c(0)$
is expected to be similar to $\lambda_{a}(T)/\lambda_{a}(0)$. This is observed in the quasi-2D organic superconductors,
where the in-plane and inter-plane penetration depth have a similar temperature dependence \cite{Carrington1999}.

The importance of the observation of a linear $T$ dependence of the superfluid density over a wide range of temperature
is that it is very difficult to produce this from extrinsic effects.  The identification by Hardy \textit{et al.}
\cite{Hardy1993} of a similar predominately linear behavior of $\lambda(T)$ in YBa$_2$Cu$_3$O$_7$ was instrumental in
identifying the $d_{x^2-y^2}$ state in the cuprates. Impurities will weaken any intrinsic linear temperature
dependence, with $\lambda(T)$ varying like $T^2$ in the dirty limit \cite{HirschfeldG93}. However, power-laws with
exponent close to two can result from, for example, a fully gapped $s_\pm$ state with impurities \cite{Chubukov08b}.
The impurities produce a non-zero zero energy density of states similar to that found in the $d$-wave case, and this
has been shown to account well for the NMR data \cite{Parker08,Chubukov08}. Hence, although power-laws (with $n\sim 2$)
point towards an unconventional gap symmetry they can be produced either with or without intrinsic nodes.  However, the
linear dependence observed here strongly points towards intrinsic nodes.

The observation of an exponential behavior of $\lambda(T)$ is also a robust feature which is not easily mimicked by
extrinsic effects, so the different behavior found here for LaFePO compared to the As based 1111 compounds detailed
above, probably points towards these compounds having intrinsically different gap symmetry / anisotropy.  Just such a
possibility was recently emphasized in the theoretical work of Graser {\it et al.} \cite{Graser08120343}.

After completion of this work we learnt of local scanning SQUID susceptibility measurements in LaFePO crystals which
also show a linear temperature dependence \cite{MolerNote}.

From precise measurements of the penetration depth in clean crystals we have shown strong evidence to suggest that the
ferro pnictide superconductor LaFePO has nodes in its superconducting gap function.  Combined with the detailed
information about the full three dimensional electronic structure of this material derived from dHvA measurements, this
should help to decide between competing theories for the origin of superconductivity in these materials.

We thank A.~Chubukov, P.~Hirschfeld and R.~Prozorov for useful comments. Work at Bristol was supported by EPSRC. Work
at Stanford was supported by the DOE, Office of Basic Energy Sciences under contract DE-AC02-76SF00515.

\end{document}